\documentclass[11pt,twoside]{atmp}

\usepackage{amsmath,amssymb}

\usepackage[all]{xy}
\usepackage{color}

\copyrightnotice{2010}{14}{1}{19}

\begin{document}

\title[Jordan derivations on $C^*$-ternary algebras]
{Jordan derivations on $C^*$-ternary algebras for a Cauchy-Jensen functional equation}

\arxurl{hep reference num}

\author[C. Park, J. M. Rassias and W.-G. Park]
{Choonkil Park$^1$, John Michael Rassias$^2$ and Won-Gil Park$^3$}

\address{$^1$ Department of Mathematics, Research Institute for Natural Sciences
Hanyang University, Seoul, 133--791, Republic of Korea\\
$^2$ Pedagogical Department, E.E., National and Capodistrian University of Athens
4, Agamemnonos Str., Aghia Paraskevi, Athens 15342, Greece\\
$^3$ Department of Mathematics Education, College of Education
Mokwon University, Daejeon 302-729, Republic of Korea}

\addressemail{$^1$baak@@hanyang.ac.kr, $^2$jrassias@@primedu.uoa.gr\\
 and $^3$wgpark@@mokwon.ac.kr}

\begin{abstract}
In this paper, we proved the generalized Hyers-Ulam stability of homomorphisms in $C^*$-
ternary algebras and of derivations on $C^*$-ternary algebras for the following Cauchy-
Jensen functional equation
$$3f\bigg(\frac{x+y+z}{3}\bigg)=2f\bigg(\frac{x+y}{2}\bigg)+f(z).$$
These were applied to investigate isomorphisms between $C^*$-ternary algebras.
\end{abstract}

\vspace{2mm}

\theoremstyle{definition} \newtheorem{df}{Definition}[section] \newtheorem{rk}[df]{Remark}
\theoremstyle{plain} \newtheorem{lem}[df]{Lemma} \newtheorem{thm}[df]{Theorem}
 \newtheorem{cor}[df]{Corollary}

\maketitle

\cutpage

\setcounter{page}{2}

\section{Introduction and preliminaries} 

Ternary structures and their generalization, the so-called $n$-ary structures, raise certain
hopes in view of their applications in physics. Some significant physical applications are as
follows (see \cite{ke97, ke00}):

(1) The algebra of {\it `nonions'} generated by two matrices
\begin{eqnarray*}
\left(\begin{array}{ccc}0&1&0\\0&0&1\\1&0&0\end{array}\right)\qquad\&\qquad
\left(\begin{array}{ccc}0&1&0\\0&0&\omega\\\omega^2&0&0\end{array}\right)\qquad
\big(\,\omega=e^{\frac{2\pi i}{3}}\,\big)
\end{eqnarray*}
was introduced by Sylvester as a ternary analog of Hamilton's quaternions (cf.
\cite{aklr}).

(2) The quark model inspired a particular brand of ternary algebraic systems. The so-called
{\it `Nambu mechanics'} is based on such structures (see \cite{dt97}).

There are also some applications, although still hypothetical, in the fractional quantum Hall
effect, the non-standard statistics, supersymmetric theory, and Yang--Baxter equation (cf.
\cite {aklr, ke00, vk96}).

A $C^*$-ternary algebra is a complex Banach space $A$, equipped with a ternary product
$(x,y,z)\mapsto[x,y,z]$ of $A^3$ into $A$, which is $\mathbb C$-linear in the outer
variables, conjugate $\mathbb C$-linear in the middle variable, and associative in the sense
that $[\,x,y,[z,w,v]\,]=[\,x,[w,z,y],v\,]=[\,[x,y,z],w,v\,]$, and satisfies
$\|\,[x,y,z]\,\|\le\|x\|\cdot\|y\|\cdot\|z\|$ and $\|\,[x, x, x]\,\|$ $=\|x\|^3$ (see
\cite{am00,ze83}). Every left Hilbert $C^*$-module is a $C^*$-ternary algebra via the ternary
product $[x,y,z]:=\langle x,y\rangle z$.

If a $C^*$-ternary algebra $(A,[\cdot,\cdot,\cdot])$ has an identity, i.e., an element
$e\in A$ such that $x=[x,e,e]=[e,e,x]$ for all $x\in A$, then it is routine to verify that
$A$, endowed with $x\circ y:=[x,e,y]$ and $x^*:=[e,x,e]$, is a unital $C^*$-algebra.
Conversely, if $(A,\circ)$ is a unital $C^*$-algebra, then $[x,y,z]:=x\circ y^*\circ z$ makes
$A$ into a $C^*$-ternary algebra.

A $\mathbb C$-linear mapping $H:A\to B$ is called a {\it $C^*$-ternary algebra homomorphism}
if
$$H([x,y,z])=[H(x),H(y),H(z)]$$
for all $x,y,z\in A$. If, in addition, the mapping $H$ is bijective, then the mapping
$H:A\to B$ is called a {\it $C^*$-ternary algebra isomorphism}. A $\mathbb C$-linear mapping
$\delta:A\to A$ is called a {\it $C^*$-ternary derivation} if
$$\delta([x,y,z])=[\delta(x),y,z]+[x,\delta(y),z]+[x,y,\delta(z)]$$
for all $x,y,z\in A$ (see \cite{am00}, \cite{mo07a}--\cite{ms06}).

In 1940, S. M. Ulam \cite{ul60} gave a talk before the Mathematics Club of the University of
Wisconsin in which he discussed a number of unsolved problems. Among these was the following
question concerning the stability of homomorphisms.

{\it We are given a group $G$ and a metric group $G'$ with metric $\rho(\cdot,\cdot)$. Given
$\epsilon>0$ , does there exist a $\delta>0$  such that if $f:G\to G'$ satisfies
$$\rho(f(xy),f(x)f(y))<\delta$$
for all $x,y\in G$, then a homomorphism $h:G\to G'$ exists with
$$\rho(f(x),h(x))<\epsilon$$
for all $x\in G$?}

In 1941, D. H. Hyers \cite{hy41} considered the case of approximately additive mappings
$f:E\to E'$, where $E$ and $E'$ are Banach spaces and $f$ satisfies {\it Hyers inequality}
$$\|f(x+y)-f(x)-f(y)\|\le\epsilon$$
for all $x,y\in E$. It was shown that the limit
$$L(x)=\lim_{n\to\infty}\frac{f(2^nx)}{2^n}$$
exists for all $x\in E$ and that $L:E\to E'$ is the unique additive mapping satisfying
$$\|f(x)-L(x)\|\le\epsilon$$
for all $x\in E$.

In 1978, $\text{Th. M. Rassias}$ \cite{ra78} provided a generalization of the D. H. Hyers'
theorem which allows the {\it Cauchy difference to be unbounded.}

\begin{thm} $(\rm\text{Th. M. Rassias})$
Let $f:E\to E'$ be a mapping from a normed vector
space $E$ into a Banach space $E'$ subject to the inequality
\begin{equation}\|f(x+y)-f(x)-f(y)\|\le\epsilon(\|x\|^p+\|y\|^p)\label{app add}
\end{equation}
for all $x,y\in E$, where $\epsilon$ and $p$ are constants with $\epsilon>0$ and $p<1$. Then
the limit
$$L(x) = \lim_{n\to\infty} \frac{f(2^n x)}{2^n}$$
exists for all $x\in E$ and $L:E\to E'$ is the unique additive mapping which
satisfies
\begin{equation}\|f(x)-L(x)\|\le\frac{2\epsilon}{2-2^p}\|x\|^p\label{add near f}
\end{equation}
for all $x\in E$. If $p<0$ then inequality $(\ref{app add})$ holds for $x,y\ne 0$ and
$(\ref{add near f})$ for $x\ne 0$.
\end{thm}

On the other hand, in 1982-1989, J. M. Rassias generalized the Hyers stability result by
presenting a weaker condition controlled by a product of different powers of norms. The
following is according to the J. M. Rassias' theorem.

\begin{thm} {\rm (J. M. Rassias)}
If it is assumed that there exist constants $\Theta\ge0$ and $p_1,p_2\in\mathbb R$ such that
$p=p_1+p_2\ne1$, and $f:E\to E'$ is a mapping from a normed space $E$ into a Banach space
$E'$ such that the inequality
$$\|f(x+y)-f(x)-f(y)\|\le\epsilon\|x\|^{p_1}\|y\|^{p_2}$$
for all $x,y\in E$, then there exists a unique additive mapping $T:E\to E'$ such that
$$\|f(x)-L(x)\|\le\frac{\Theta}{2-2^p}\|x\|^p$$
for all $x\in E$.
\end{thm}

In 1990, $\text{Th. M. Rassias}$ \cite{ra90} during the 27$^{\,\rm\text th}$ International
Symposium on Functional Equations asked the question whether such a theorem can also be
proved for $p\ge1$. In 1991, Z. Gajda \cite{gj91} following the same approach as in
$\text{Th. M. Rassias}$ \cite{ra78}, gave an affirmative solution to this question for $p>1$.
It was shown by Z. Gajda \cite{gj91}, as well as by $\text{Th. M. Rassias}$ and P. \v{S}emrl
\cite{rs92} that one cannot prove a Th. M. Rassias' type theorem when $p=1$. The
counterexamples of Z. Gajda \cite{gj91}, as well as of $\text{Th. M. Rassias}$ and P.
\v{S}emrl \cite{rs92} have stimulated several mathematicians to invent new definitions of
{\it approximately additive} or {\it approximately linear} mappings, cf. P. G\u{a}vruta
\cite{ga94}, S.-M. Jung \cite{jg96}, who among others studied the Hyers-Ulam stability of
functional equations. The inequality (\ref{app add}) that was introduced for the first time
by $\text{Th. M. Rassias}$ \cite{ra78} provided a lot of influence in the development of a
generalization of the Hyers-Ulam stability concept. This new concept is known as {\it
generalized Hyers-Ulam stability} of functional equations (cf. the books of P. Czerwik
\cite{cz02}, D. H. Hyers et al. \cite{hir98}).

P. G\u{a}vruta \cite{ga94} provided a further generalization of $\text{Th. M. Rassias}$'
Theorem. In 1996, G. Isac and $\text{Th. M. Rassias}$ \cite{ir96} applied the generalized
Hyers-Ulam stability theory to prove fixed point theorems and study some new applications in
Nonlinear Analysis. In \cite{hir08}, D. H. Hyers et al. studied
the asymptoticity aspect of Hyers-Ulam stability of mappings. Several papers have been
published on various generalizations and applications of Hyers-Ulam stability and generalized
Hyers-Ulam stability to a number of functional equations and mappings, for example: quadratic
functional equation, invariant means, multiplicative mappings - superstability, bounded $n$th
differences, convex functions, generalized orthogonality functional equation, Euler-Lagrange
functional equation introduced by J. M. Rassias in 1992-1998, Navier-Stokes equations.
Several mathematician have contributed works on these subjects (see \cite{bm05},
\cite{npl09}--\cite{sk83}).

In Section 2, we prove the generalized Hyers-Ulam stability of homomorphisms in $C^*$-ternary
algebras for the Cauchy-Jensen additive mappings.

In Section 3, we investigate isomorphisms between unital $C^*$-ternary algebras associated
with the Cauchy-Jensen additive mappings.

In Section 4, we prove the generalized Hyers-Ulam stability of derivations on $C^*$-ternary
algebras for the Cauchy-Jensen additive mappings.

\section{Stability of homomorphisms in $C^*$-ternary algebras} 

Throughout this section, assume that $A$ is a $C^*$-ternary algebra with norm $\|\cdot\|_A$
and that $B$ is a $C^*$-ternary algebra with norm $\|\cdot\|_B$.

For a given mapping $f:A\to B$, we define
$$D_\mu f(x,y,z):=3f\Big(\frac{\mu x+\mu y+\mu z}{3}\Big)-2\mu f\Big(\frac{x+y}{2}\Big)
-\mu f(z)$$
for all $\mu\in{\mathbb T}^1:=\{\lambda\in{\mathbb C}\,\mid \,|\lambda|=1\}$ and all
$x,y,z\in A$.

\begin{lem} 
Let $f:A\to B$ be a mapping such that
\begin{equation}D_\mu f(x,y,z)=0\label{add}\end{equation}
\noindent for all $\mu\in{\mathbb T}^1$ and all $x,y,z\in A$.
Then $f$ is $\mathbb C$-linear.
\end{lem}

\noindent{\it Proof.}
Letting $\mu=-1$ and $x=y=z=0$ in (\ref{add}), we gain $f(0)=0$. Putting $\mu=1$, $y=-x$ and
$z=2x$ in (\ref{add}), we get $3f(\frac{2}{3}x)=f(2x)$ for all $x\in A$. So we have
$3f(x)=f(3x)$ for all $x\in A$. Setting $\mu=1$, $x=0$ in (\ref{add}), we gain
$$3f\Big(\frac{y+z}{3}\Big)=2f\Big(\frac{y}{2}\Big)+f(z)$$
for all $y,z\in A$. So we get $f(y+z)=2f\big(\frac{y}{2}\big)+f(z)$ for all
$y,z\in A$. Taking $z=0$ in the above equation, we have $f(y)=2f\big(\frac{y}{2}\big)$ for
all $y\in A$. Thus we obtain that $f(y+z)=f(y)+f(z)$ for all $y,z\in A$. Hence $f$ is
additive.

Letting $y=z=0$ in (\ref{add}), we gain
$3f\big(\frac{\mu}{3}x\big)=2\mu f\big(\frac{x}{2}\big)$ for all $\mu\in{\mathbb T}^1$ and
all $x\in A$. Since $f$ is additive, $f(\mu x)=f\big(3\,\frac{\mu}{3}x\big)
=3f\big(\frac{\mu}{3}x\big)=2\mu f\big(\frac{x}{2}\big)=\mu f\big(2\,\frac{x}{2}\big)
=\mu f(x)$ for all $\mu\in{\mathbb T}^1$ and all $x\in A$. Now let $\lambda\in\mathbb C$ and
$M$ an integer greater than $2|\lambda|$. Since $\big|\frac{\lambda}{M}\big|<\frac{1}{2}$,
there is $t\in\big(\frac{\pi}{3},\frac{\pi}{2}\big]$ such that $\big|\frac{\lambda}{M}\big|
=\cos t=\frac{e^{it}+e^{-it}}{2}.$ Now $\frac{\lambda}{M}=\big|\frac{\lambda}{M}\big|\mu$ for
some $\mu\in{\mathbb T}^1$. Thus we have
\begin{eqnarray*}
&&f(\lambda x)=f\bigg(M\frac{\lambda}{M}x\bigg)=Mf\bigg(\frac{\lambda}{M}x\bigg)
 =Mf\bigg(\bigg|\frac{\lambda}{M}\bigg|\mu x\bigg)\\
&&\qquad\ \ \,=Mf\bigg(\frac{e^{it}+e^{-it}}{2}\mu x\bigg)=\frac{1}{2}Mf\big(e^{it}\mu x+e^{-it}\mu x\big)\\
&&\qquad\ \ \,=\frac{1}{2}M\big[e^{it}\mu f(x)+e^{-it}\mu f(x)\big]=\lambda f(x)
\end{eqnarray*}
for all $x\in A$. So the mapping $f:A\to B$ is $\mathbb C$-linear.
\qed

We prove the generalized  Hyers-Ulam stability of homomorphisms in $C^*$-ternary algebras for
the functional equation $D_{\mu}f(x,y,z)=0$.

\begin{thm} 
Let $r>3$ and $\theta$ be positive real numbers, and let $f:A\to B$ be a mapping satisfying
$f(0)=0$ such that
\begin{eqnarray}
\|D_\mu f(x,y,z)\|_B &\le& \theta(\|x\|^r_A+\|y\|^r_A+\|z\|^r_A),\label{app add r>3}\\
\|f([x,y,z])-[f(x),f(y),f(z)]\|_B &\le& \theta(\|x\|^r_A+\|y\|^r_A+\|z\|^r_A)\label{app ter
r>3}
\end{eqnarray}
for all $\mu\in{\mathbb T}^1$ and all $x,y,z\in A$.
Then there exists a unique $C^*$-ternary algebra homomorphism $H:A\to B$ such that
\begin{equation}
\|f(x)-H(x)\|_B\ \le\ \theta\,\frac{3^r+2}{3^r-3}\,\|x\|^r_A\label{homo near f r>3}
\end{equation}
for all $x\in A$.
\end{thm}

\noindent{\it Proof.}
Letting $\mu=1$ and $y=-x$ and $z=3x$ in (\ref{app add r>3}), we obtain
\begin{equation}
\|3f(x)-f(3x)\|_B\le\theta(2+3^r)\|x\|^r_A\label{3x r>3}
\end{equation}
for all $x\in A$. So we get
$$\Big\|f(x)-3f\Big(\frac{x}{3}\Big)\Big\|_B\le\theta\Big(\frac{2}{3^r}+1\Big)\|x\|^r_A$$
for all $x\in A$. Thus we have
\begin{eqnarray}
&& \Big\|3^lf\Big(\frac{x}{3^l}\Big)-3^mf\Big(\frac{x}{3^m}\Big)\Big\|_B
 \le\sum_{j=l}^{m-1}\Big\|3^jf\Big(\frac{x}{3^j}\Big)
 -3^{j+1}f\Big(\frac{x}{3^{j+1}}\Big)\Big\|_B\nonumber\\
&& \qquad \le\theta\,\bigg(\frac{2}{3^r}+1\bigg)\sum_{j=l}^{m-1}3^{j(1-r)}\|x\|_A^r
 \,=\,\theta\,\frac{3^r+2}{3^r-3}\,\big[3^{l(1-r)}-3^{m(1-r)}\big]\|x\|_A^r\label{Cauchy r>3}
\end{eqnarray}
for all nonnegative integers $m$ and $l$ with $m>l$ and all $x\in A$. It follows
from (\ref{Cauchy r>3}) that the sequence $\{3^nf(\frac{x}{3^n})\}$ is a Cauchy sequence for
all $x\in A$. Since $B$ is complete, the sequence $\{3^nf(\frac{x}{3^n})\}$ converges for all
$x\in A$. Hence one can define a mapping $H:A\to B$ by
$$H(x):=\lim_{n\to\infty}3^nf\Big(\frac{x}{3^n}\Big)$$
for all $x\in A$. Moreover, letting $l=0$ and passing the limit $m\to\infty$ in
(\ref{Cauchy r>3}), we get (\ref{homo near f r>3}).

It follows from (\ref{app add r>3}) that
\begin{eqnarray*}
&& \Big\|3H\Big(\frac{x+y+z}{3}\Big)-2H\Big(\frac{x+y}{2}\Big)-H(z)\Big\|_B\\
&& \qquad=\lim_{n\to\infty}3^n\Big\|\,3f\Big(\frac{x+y+z}{3^{n+1}}\Big)
 -2f\Big(\frac{x+y}{2\cdot 3^n}\Big)-f\Big(\frac{z}{3^n}\Big)\Big\|_B\\
&& \qquad\le\lim_{n\to\infty}3^{n(1-r)}\theta(\|x\|^r_A+\|y\|^r_A+\|z\|^r_A)=0
\end{eqnarray*}
for all $x,y,z\in A$. So we get
$$3H\Big(\frac{x+y+z}{3}\Big)=2H\Big(\frac{x+y}{2}\Big)+H(z)$$
for all $x,y,z\in A$. Since $f(0)=0$, by the same methods as in proof of Lemma 2.1,
the mapping $H:A\to B$ is additive.

By the same reasoning as in the proof of Theorem 2.1 in \cite{pa05b}, the mapping $H:A\to$
$B$ is $\mathbb C$-linear.
It follows from (\ref{app ter r>3}) and (\ref{Cauchy r>3}) that
\begin{eqnarray*}
&& \|H([x,y,z])-[H(x),H(y),H(z)]\|_B\\
&& \qquad=\lim_{n\to\infty}\Big\|\,3^nf\Big(\frac{1}{3^n}[x,y,z]\Big)-\Big[\,3^n
 f\Big(\frac{x}{3^n}\Big),3^nf\Big(\frac{y}{3^n}\Big),3^nf\Big(\frac{z}{3^n}\Big)\Big]
 \Big\|_B\\
&& \qquad=\lim_{n\to\infty}\Big[\,\Big\|\,3^nf\Big(\frac{1}{3^n}[x,y,z]\Big)-3^{2n}
 f\Big(\frac{1}{3^{2n}}[x,y,z]\Big)\Big\|_B\\
&& \qquad\qquad\quad+\Big\|\,3^{2n}f\Big(\frac{1}{3^{2n}}[x,y,z]\Big)-3^{3n}
 f\Big(\frac{1}{3^{3n}} [x,y,z]\Big)\Big\|_B\\
&& \qquad\qquad\quad+\Big\|\,3^{3n}f\Big(\Big[\frac{x}{3^n},\frac{y}{3^n},\frac{z}{3^n}\Big]
 \Big)\\
&& \qquad\qquad\quad-\Big[3^nf\Big(\frac{x}{3^n}\Big),3^nf\Big(\frac{y}{3^n}\Big),3^nf\Big(\frac{z}{3^n}
 \Big)\Big]\Big\|_B\,\Big]
\end{eqnarray*}\begin{eqnarray*}
&& \qquad\le\lim_{n\to\infty}\bigg[\,\theta\Big(\frac{2}{3^r}+1\Big)
 \sum_{j=n}^{2n-1}3^{j(1-r)}\|[x,y,z]\|_A^r\\
&& \qquad\qquad\quad+\theta\Big(\frac{2}{3^r}+1\Big)\sum_{j=2n}^{3n-1}
 3^{j(1-r)}\,\|[x,y,z]\|_A^r\\
&& \qquad\qquad\quad+\,3^{n(3-r)}\theta(\|x\|^r_A+\|y\|^r_A+\|z\|^r_A)\bigg]\\
&& \qquad=\theta\,\frac{3^r+2}{3^r-3}\,\|[x,y,z]\|_A^r\,\lim_{n\to\infty}\big[3^{n(1-r)}
 -3^{3n(1-r)}\big]\\
&& \qquad\qquad\quad+\,\theta(\|x\|^r_A+\|y\|^r_A+\|z\|^r_A)\lim_{n\to\infty}3^{n(3-r)}\\
&& \qquad=0
\end{eqnarray*}
for all $x,y,z\in A$. So
$$H([x,y,z])=[H(x),H(y),H(z)]$$
for all $x,y,z\in A$.

Now, let $T:A\to B$ be another additive mapping satisfying (\ref{homo near f r>3}). Then we
have
\begin{eqnarray*}
&& \|H(x)-T(x)\|_B\,=\,3^n\Big\|H\Big(\frac{x}{3^n}\Big)-T\Big(\frac{x}{3^n}\Big)\Big\|_B\\
&& \qquad\le3^n\Big[\,\Big\|H\Big(\frac{x}{3^n}\Big)-f\Big(\frac{x}{3^n}\Big)\Big\|_B
 +\Big\|f\Big(\frac{x}{3^n}\Big)-T\Big(\frac{x}{3^n}\Big)\Big\|_B\Big]\\
&& \qquad\le\frac{2\theta}{3^{n(r-1)}}\,\frac{3^r+2}{3^r-3}\,\|x\|^r_A,
\end{eqnarray*}
which tends to zero as $n\to\infty$ for all $x\in A$. So we can conclude that
$H(x)=T(x)$ for all $x\in A$. This proves the uniqueness of $H$. Thus the mapping $H:A\to B$
is a unique $C^*$-ternary algebra homomorphism satisfying (\ref{homo near f r>3}).
\qed

J. M. Rassias presents the following counterexample modified by the well-known
counterexample of Z. Gajda \cite{gj91} for the following Cauchy-Jensen functional
equation:
$$3f\Big(\frac{x+y+z}{3}\Big)=2f\Big(\frac{x+y}{2}\Big)+f(z).$$
Fix $\theta>0$ and put $\mu:=\frac{\theta}{6}$. Define a function
$f:\mathbb{R}\to\mathbb{R}$ given by
$$f(x):=\sum_{n=0}^\infty\frac{\phi(2^nx)}{2^n}$$
for all $x\in\mathbb{R}$, where
$$\phi(x):=
\begin{cases}
\mu  &\text{if}\quad x\ge1\\
\mu x&\text{if}\ \ -1<x<1\\
-\mu &\text{if}\quad x\le-1
\end{cases}$$
for all $x\in\mathbb{R}$. It was proven in \cite{gj91} that
$$|f(x+y)-f(x)-f(y)|\le\theta(|x|+|y|)$$
for all $x,y\in\mathbb{R}$. From the above inequality, one can obtain that
\begin{eqnarray}
&&|f(x+y+z)-f(x)-f(y)-f(z)|\label{3C}\\
&&\qquad\le\frac{1}{3}\,\big[\,|f(x+y+z)-f(x+y)-f(z)|\nonumber\\
&&\qquad\qquad+|f(x+y+z)-f(x+z)-f(y)|\nonumber\\
&&\qquad\qquad+|f(x+y+z)-f(y+z)-f(x)|\nonumber\\
&&\qquad\qquad+|f(x+y)-f(x)-f(y)|+|f(x+z)-f(x)-f(z)|\nonumber\\
&&\qquad\qquad+|f(y+z)-f(y)-f(z)|\,\big]\nonumber\\
&&\qquad\le\frac{5}{3}\theta(|x|+|y|+|z|)\nonumber
\end{eqnarray}
and
\begin{eqnarray}
&&\Big|2f\Big(\frac{x+y}{2}\Big)-f(x)-f(y)\Big|\label{2J}\\
&&\qquad\le2\Big|f\Big(\frac{x}{2}+\frac{y}{2}\Big)-f\Big(\frac{x}{2}\Big)
 -f\Big(\frac{y}{2}\Big)\Big|\nonumber\\
&&\qquad\quad+\bigg|-\Big[f\Big(\frac{x}{2}+\frac{x}{2}\Big)-f\Big(\frac{x}{2}\Big)
 -f\Big(\frac{x}{2}\Big)\Big]\bigg|\nonumber\\
&&\qquad\quad+\bigg|-\Big[f\Big(\frac{y}{2}+\frac{y}{2}\Big)-f\Big(\frac{y}{2}\Big)
 -f\Big(\frac{y}{2}\Big)\Big]\bigg|\nonumber\\
&&\qquad\le2\theta(|x|+|y|)\nonumber
\end{eqnarray}
for all $x,y,z\in\mathbb{R}$. By the inequality (\ref{3C}), we see that
\begin{eqnarray}
&&\Big|3f\Big(\frac{x+y+z}{3}\Big)-f(x)-f(y)-f(z)\Big|\label{3J}
\end{eqnarray}\begin{eqnarray*}
&&\qquad\le3\Big|f\Big(\frac{x}{3}+\frac{y}{3}+\frac{z}{3}\Big)-f\Big(\frac{x}{3}\Big)
 -f\Big(\frac{y}{3}\Big)-f\Big(\frac{z}{3}\Big)\Big|\nonumber\\
&&\qquad\quad+\bigg|-\Big[f\Big(\frac{x}{3}+\frac{x}{3}+\frac{x}{3}\Big)-f\Big(\frac{x}{3}
 \Big)-f\Big(\frac{x}{3}\Big)-f\Big(\frac{x}{3}\Big)\Big]\bigg|\nonumber\\
&&\qquad\quad+\bigg|-\Big[f\Big(\frac{y}{3}+\frac{y}{3}+\frac{y}{3}\Big)\,-f\Big(\frac{y}{3}
 \Big)-f\Big(\frac{y}{3}\Big)-f\Big(\frac{y}{3}\Big)\Big]\bigg|\nonumber\\
&&\qquad\quad+\bigg|-\Big[f\Big(\frac{z}{3}+\frac{z}{3}+\frac{z}{3}\Big)\,-f\Big(\frac{z}{3}
 \Big)-f\Big(\frac{z}{3}\Big)-f\Big(\frac{z}{3}\Big)\Big]\bigg|\nonumber\\
&&\qquad\le\frac{10}{3}\theta(|x|+|y|+|z|)\nonumber
\end{eqnarray*}
for all $x,y,z\in\mathbb{R}$.
From the inequalities (\ref{2J}) and (\ref{3J}), we obtain that
\begin{eqnarray*}
&&\Big|3f\Big(\frac{x+y+z}{3}\Big)-2f\Big(\frac{x+y}{2}\Big)-f(z)\Big|\\
&&\qquad\le\Big|3f\Big(\frac{x+y+z}{3}\Big)-f(x)-f(y)-f(z)\Big|\\
&&\qquad\quad+\Big|-\Big[2f\Big(\frac{x+y}{2}\Big)-f(x)-f(y)\Big]\Big|\\
&&\qquad\le\frac{2}{3}\theta(8|x|+8|y|+5|z|)\le\frac{16}{3}\theta(|x|+|y|+|z|)
\end{eqnarray*}
for all $x,y,z\in\mathbb{R}$. But we observe from \cite{gj91} that
$$\frac{f(x)}{x}\to\infty\quad\text{as}\quad x\to\infty$$
and so
$$\frac{|f(x)-g(x)|}{|x|}\ (x\ne0)\ \text{is unbounded,}$$
where $g:\mathbb{R}\to\mathbb{R}$ is the function given by
$$g(x):=\lim_{n\to\infty}3^nf\Big(\frac{x}{3^n}\Big)$$
for all $x\in\mathbb{R}$. Thus the function $f$ provides an example to the effect
that Theorem 2.2 fails to hold for $r=1$.

\begin{thm}
Let $r<1$ and $\theta$ be positive real numbers, and let $f:A\to B$ be a mapping satisfying
$(\ref{app add r>3})$, $(\ref{app ter r>3})$ and $f(0)=0$. Then there exists a unique $C^*$
-ternary algebra homomorphism $H:A\to B$ such that
\begin{equation}\|f(x)-H(x)\|_B\le\theta\,\frac{\,2+3^r}{\,3-3^r}\,\|x\|_A^r
\label{homo near f r<1}\end{equation}
for all $x\in A$.
\end{thm}

\noindent{\it Proof.}
It follows from (\ref{3x r>3}) that
$$\bigg\|f(x)-\frac{1}{3}f(3x)\bigg\|_B \le \theta\,\frac{\,2+3^r}{3}\,\|x\|^r_A$$
for all $x \in A$. So
\begin{eqnarray}
&& \bigg\|\frac{1}{3^l}f(3^l x)-\frac{1}{3^m}f(3^mx)\bigg\|_B
 \le\sum_{j=l}^{m-1}\bigg\|\frac{1}{3^j}f(3^jx)-\frac{1}{3^{j+1}}f(3^{j+1}x)\bigg\|_B
 \nonumber\\
&& \qquad\le\theta\,\frac{\,2+3^r}{3}\sum_{j=l}^{m-1}3^{j(r-1)}\|x\|_A^r
 \,=\,\theta\,\frac{2+3^r}{3-3^r}\big[3^{l(r-1)}-3^{m(r-1)}\big]\|x\|_A^r
 \label{Cauchy r<1}
\end{eqnarray}
for all nonnegative integers $m$ and $l$ with $m>l$ and all $x\in A$. It follows
from (\ref{Cauchy r<1}) that the sequence $\{\frac{1}{3^n} f(3^nx)\}$ is a Cauchy sequence
for all $x\in A$. Since $B$ is complete, the sequence $\{\frac{1}{3^n} f(3^nx)\}$ converges
for all $x\in A$. So one can define the mapping $H:A\to B$ by
$$H(x):=\lim_{n\to\infty}\frac{1}{3^n}f(3^nx)$$
for all $x\in A$. Moreover, letting $l=0$ and passing the limit $m\to\infty$ in
(\ref{Cauchy r<1}), we get (\ref{homo near f r<1}).

By similar arguments to the proof of Theorem 2.2, the mapping $H:A\to B$ is $\mathbb C$
-linear. It follows from (\ref{app ter r>3}) and (\ref{Cauchy r<1}) that
\begin{eqnarray*}
&& \|H([x,y,z])-[H(x),H(y),H(z)]\|_B\\
&& \qquad\le\theta\,\frac{2+3^r}{3-3^r}\,\|[x,y,z]\|_A^r\,\lim_{n\to\infty}\big[3^{n(r-1)}
 -3^{3n(r-1)}\big]\\
&& \qquad\qquad\quad+\,\theta(\|x\|^r_A+\|y\|^r_A+\|z\|^r_A)\lim_{n\to\infty}3^{n(r-3)}\\
&& \qquad=0
\end{eqnarray*}
for all $x,y,z\in A$. So
$$H([x,y,z])=[H(x),H(y),H(z)]$$
for all $x,y,z\in A$. Now, let $T:A\to B$ be another additive mapping satisfying
(\ref{homo near f r<1}). Then we have
$$\|H(x)-T(x)\|_B\le2\,\theta\,3^{n(r-1)}\,\frac{2+3^r}{3-3^r}\,\|x\|^r_A,$$
which tends to zero as $n\to\infty$ for all $x\in A$. So we can conclude that $H(x)=T(x)$ for
all $x\in A$. This proves the uniqueness of $H$. Thus the mapping $H:A\to B$ is a unique
$C^*$-ternary algebra homomorphism satisfying (\ref{homo near f r<1}).
\qed

\begin{thm} 
Let $r>\frac{1}{3}$ and $\theta$ be positive real numbers, and let $f:A\to B$ be a
mapping satisfying $f(0)=0$ such that
\begin{eqnarray}
\|D_{\mu}f(x,y,z)\|_B  &\le& \theta\cdot\|x\|^{r}_A\cdot\|y\|^{r}_A\cdot\|z\|^{r}_A,
\label{app add r>1/3}\\
\|f([x,y,z])-[f(x),f(y),f(z)]\|_B  &\le& \theta\cdot\|x\|^r_A\cdot\|y\|^r_A\cdot\|z\|^r_A
\label{app ter r>1/3}
\end{eqnarray}
for all $\mu\in{\mathbb T}^1$ and all $x,y,z\in A$. Then there exists a unique
$C^*$-ternary algebra
homomorphism $H:A\to B$ such that
\begin{equation}\|f(x)-H(x)\|_B\le\frac{3^r\theta}{27^r-3}\|x\|^{3r}_A
\label{homo near f r>1/3}
\end{equation}
for all $x\in A$.
\end{thm}

\noindent{\it Proof.}
Letting $\mu=1$ and $y=-x$ and $z=3x$ in (\ref{app add r>1/3}), we get
\begin{equation}\|f(3x)-3 f(x)\|_B\le3^r\theta\|x\|^{3r}_A\label{3x r>1/3}\end{equation}
for all $x\in A$. So
$$\Big\|f(x)-3f\Big(\frac{x}{3}\Big)\Big\|_B\le\frac{\theta}{9^r}\|x\|^{3r}_A$$
for all $x \in A$. Hence
\begin{eqnarray}
&& \Big\|3^lf\Big(\frac{x}{3^l}\Big)-3^m f\Big(\frac{x}{3^m}\Big)\Big\|_B
 \le\sum_{j=l}^{m-1} \Big\|3^jf\Big(\frac{x}{3^j}\Big)
 -3^{j+1}f\Big(\frac{x}{3^{j+1}}\Big)\Big\|_B\nonumber\\
&& \qquad\le\frac{\theta}{9^r}\sum_{j=l}^{m-1}3^{j(1-3r)}\|x\|_A^{3r}
 \,=\,\frac{\theta}{9^r-3^{1-r}}\big[3^{l(1-3r)}-3^{m(1-3r)}\big]\|x\|_A^{3r}
 \label{Cauchy r>1/3}
\end{eqnarray}
for all nonnegative integers $m$ and $l$ with $m>l$ and all $x\in A$.
It follows from (\ref{Cauchy r>1/3}) that the sequence $\{3^nf(\frac{x}{3^n})\}$ is a Cauchy
sequence for all $x\in A$. Since $B$ is complete,
the sequence $\{3^nf(\frac{x}{3^n})\}$
converges. So one can define the mapping $H:A\to B$ by
$$H(x):=\lim_{n\to\infty}3^n f\Big(\frac{x}{3^n}\Big)$$
for all $x\in A$. Moreover, letting $l=0$ and passing the limit $m\to\infty$ in
(\ref{Cauchy r>1/3}), we get (\ref{homo near f r>1/3}).

The rest of the proof is similar to the proof of Theorem 2.2.
\qed

\begin{thm} 
Let $r<\frac{1}{3}$ and $\theta$ be positive real numbers, and let $f:A\to B$ be a mapping
satisfying $(\ref{app add r>1/3})$, $(\ref{app ter r>1/3})$ and $f(0)=0$. Then there exists a
unique $C^*$-ternary
algebra homomorphism $H:A\to B$ such that
\begin{equation}\|f(x)-H(x)\|_B\le\frac{3^r\theta}{3-27^r}\|x\|^{3r}_A
\label{homo near f r<1/3}\end{equation}
for all $x\in A$.
\end{thm}

\noindent{\it Proof.}
It follows from (\ref{3x r>1/3}) that
$$\bigg\|f(x)-\frac{1}{3}f(3x)\bigg\|_B\le3^{r-1}\theta\|x\|^{3r}_A$$
for all $x\in A$. So
\begin{eqnarray}
&& \bigg\|\frac{1}{3^l}f(3^lx)-\frac{1}{3^m}f(3^m x)\bigg\|_B
 \le\sum_{j=l}^{m-1}\bigg\|\frac{1}{3^j}f(3^jx)-\frac{1}{3^{j+1}}f(3^{j+1}x)\bigg\|_B
 \nonumber\\
&& \qquad\le3^{r-1}\theta\sum_{j=l}^{m-1}3^{j(3r-1)}\|x\|_A^{3r}
 \,=\,\frac{3^{r-1}\theta}{1-3^{3r-1}}\big[3^{l(3r-1)}-3^{m(3r-1)}\big]\|x\|_A^{3r}
 \label{Cauchy r<1/3}
\end{eqnarray}
for all nonnegative integers $m$ and $l$ with $m>l$ and all $x\in A$. It follows
from (\ref{Cauchy r<1/3}) that the sequence $\{\frac{1}{3^n}f(3^nx)\}$ is a Cauchy sequence
for all $x\in A$. Since $B$ is complete, the sequence $\{\frac{1}{3^n}f(3^nx)\}$ converges
for all $x\in A$. So one can define the mapping $H:A\to B$ by
$$H(x):=\lim_{n\to\infty}\frac{1}{3^n}f(3^nx)$$
for all $x\in A$. Moreover, letting $l=0$ and passing the limit $m\to\infty$ in
(\ref{Cauchy r<1/3}), we get (\ref{homo near f r<1/3}).

The rest of the proof is similar to the proof of Theorem 2.2.
\qed

\section{Isomorphisms between $C^*$-ternary algebras} 

Throughout this section, assume that $A$ is a unital $C^*$-ternary algebra with norm
$\|\cdot\|_A$ and unit $e$, and that $B$ is a unital $C^*$-ternary algebra with norm
$\|\cdot\|_B$ and unit $e'$.

We investigate isomorphisms between $C^*$-ternary algebras associated with the functional
equation $D_{\mu}f(x,y,z)=0$.

\begin{thm} 
Let $r>1$ and $\theta$ be positive real numbers, and let $f:A\to B$ be a
bijective mapping satisfying $(\ref{app add r>3})$ and $f(0)=0$ such that
\begin{equation}f([x,y,z])\,=\,[f(x),f(y),f(z)]\label{ter}\end{equation}
for all $x,y,z\in A$. If $\operatorname{lim}_{n\to\infty}3^nf(\frac{e}{3^n})=e'$,
then the mapping $f:A\to B$ is a $C^*$-ternary algebra isomorphism.
\end{thm}

\noindent{\it Proof.}
By the same argument as in the proof of Theorem 2.2, one can obtain a $\mathbb C$-linear
mapping $H:A\to B$ satisfying (\ref{homo near f r>3}). The mapping $H$ is given by
$$H(x):=\lim_{n\to\infty}3^nf\Big(\frac{x}{3^n}\Big)$$
for all $x\in A$.

Since $f([x,y,z])=[f(x),f(y),f(z)]$ for all $x,y,z\in A$,
\begin{eqnarray*}
H([x,y,z]) &=& \lim_{n\to\infty}3^{3n}f\bigg(\frac{1}{3^{3n}}[x,y,z]\bigg)=\lim_{n\to\infty}
 3^{3n}f\Big(\Big[\frac{x}{3^n},\frac{y}{3^n},\frac{z}{3^n}\Big]\Big)\\
&=& \lim_{n\to\infty}\Big[3^nf\Big(\frac{x}{3^n}\Big),3^nf\Big(\frac{y}{3^n}\Big),3^n
 f\Big(\frac{z}{3^n}\Big)\Big]\\
&=& [H(x),H(y),H(z)]
\end{eqnarray*}
for all $x,y,z\in A$.
So the mapping $H:A\to B$ is a $C^*$-ternary algebra homomorphism.

It follows from (\ref{ter}) that
\begin{eqnarray*}
H(x) &=& H([e,e,x])=\lim_{n\to\infty}3^{2n}f\bigg(\frac{1}{3^{2n}}[e,e,x]\bigg)
 =\lim_{n\to\infty}3^{2n}f\Big(\Big[\frac{e}{3^n},\frac{e}{3^n},x\Big]\Big)\\
&=& \lim_{n\to\infty}\Big[3^nf\Big(\frac{e}{3^n}\Big),3^nf\Big(\frac{e}{3^n}\Big),f(x)\Big]
 =[e',e',f(x)]=f(x)
\end{eqnarray*}
for all $x \in A$. Hence the bijective mapping $f:A\to B$ is a $C^*$-ternary
algebra isomorphism.
\qed

\begin{thm}  
Let $r<1$ and $\theta$ be positive real numbers, and let $f:A\to B$ be a bijective mapping
satisfying $(\ref{app add r>3})$, $(\ref{ter})$ and $f(0)=0$. If
$\lim_{n\to\infty}\frac{1}{3^n}f(3^ne)=e'$, then the mapping $f:A\to B$ is a $C^*$-ternary
algebra isomorphism.
\end{thm}

\noindent{\it Proof.}
By the same argument as in the proof of Theorem 2.3, one can obtain a $\mathbb C$-linear
mapping $H:A\to B$ satisfying (\ref{homo near f r<1}).

The rest of the proof is similar to the proof of Theorem 3.1.
\qed

\begin{thm} 
Let $r>\frac{1}{3}$ and $\theta$ be positive real numbers, and let $f:A\to B$ be a bijective
mapping satisfying $(\ref{app add r>1/3})$, $(\ref{ter})$ and $f(0)=0$. If
$\,\lim_{n\to\infty}3^nf(\frac{e}{3^n})=e'$, then the mapping $f:A\to B$ is a
$C^*$-ternary algebra isomorphism.
\end{thm}

\noindent{\it Proof.}
By the same argument as in the proof of Theorem 2.4, one can obtain a $\mathbb C$-linear
mapping $H:A\to B$ satisfying (\ref{homo near f r>1/3}).

The rest of the proof is similar to the proof of Theorem 3.1.
\qed

\begin{thm} 
Let $r<\frac{1}{3}$ and $\theta$ be positive real numbers, and let $f:A\to B$ be a bijective
mapping satisfying $(\ref{app add r>1/3})$, $(\ref{ter})$ and $f(0)=0$. If
$\lim_{n\to\infty}\frac{1}{3^n}f(3^ne)=e'$, then the mapping $f:A\to B$ is a $C^*$-ternary
algebra isomorphism.
\end{thm}

\noindent{\it Proof.}
By the same argument as in the proof of Theorem 2.5, one can obtain a $\mathbb{C}$-linear
mapping $H:A\to B$ satisfying (\ref{homo near f r<1/3}).

The rest of the proof is similar to the proof of Theorem 3.1.
\qed

\section{Stability of $C^*$-ternary derivations on $C^*$-ternary algebras} 

Throughout this section, assume that $A$ is a $C^*$-ternary algebra with norm $\|\cdot\|_A$.

We prove the generalized Hyers-Ulam stability of $C^*$-ternary derivations on
$C^*$-ternary algebras for the functional equation $D_{\mu}f(x,y,z)=0$.

\begin{thm} 
Let $r>3$ and $\theta$ be positive real numbers, and let $f:A\to A$ be a
mapping satisfying
$f(0)=0$ such that
\begin{eqnarray}
&& \|D_\mu f(x,y,z)\|_A\,\le\,\theta(\|x\|^r_A+\|y\|^r_A+\|z\|^r_A),
\label{app add r>3 B=A}\\
&& \|f([x,y,z])-[f(x),y,z]-[x,f(y),z]-[x,y,f(z)]\|_A\nonumber\\
&& \qquad\qquad\qquad\quad\,\le\,\theta\,(\|x\|^r_A+\|y\|^r_A+\|z\|^r_A)
\label{app ter deri r>3}
\end{eqnarray}
for all $\mu\in{\mathbb T}^1$ and all $x,y,z\in A$. Then there exists a unique
$C^*$-ternary derivation
$\delta:A\to A$ such that
\begin{equation}\|f(x)-\delta(x)\|_A\ \le\ \theta\,\frac{3^r+2}{3^r-3}\,\|x\|^r_A
\label{ter deri near f r>3}\end{equation}
for all $x\in A$.
\end{thm}

\noindent{\it Proof.}
By the same argument as in the proof of Theorem 2.2, one can obtain a $\mathbb C$-linear
mapping $\delta:A\to B$ satisfying (\ref{ter deri near f r>3}). The mapping
$\delta$ is given by
$$\delta(x):=\lim_{n\to\infty}3^nf\Big(\frac{x}{3^n}\Big)$$
for all $x\in A$.

By the same reasoning as in the proof of Theorem 2.1 of \cite{pa05b}, the mapping
$\delta:A\to A$ is $\mathbb C$-linear.

It follows from (\ref{app ter deri r>3}) that
\begin{eqnarray*}
&& \|\delta([x,y,z])-[\delta(x),y,z]-[x,\delta(y),z]-[x,y,\delta(z)]\|_A\\
&& =\lim_{n\to\infty}\bigg\|3^{3n}f\bigg(\frac{[x,y,z]}{3^{3n}}\bigg)-3^{2n}\Big[3^n
 f\Big(\frac{x}{3^n}\Big),\frac{y}{3^n},\frac{z}{3^n}\Big]\\
&& \qquad\quad\,-3^{2n}\Big[\frac{x}{3^n},3^n
 f\Big(\frac{y}{3^n}\Big),\frac{z}{3^n}\Big]
 -3^{2n}\Big[\frac{x}{3^n},\frac{y}{3^n},3^nf\Big(\frac{z}{3^n}\Big)\Big]
 \bigg\|_A\\
&& \le\lim_{n\to\infty}3^{n(3-r)}\theta(\|x\|^r_A+\|y\|^r_A+\|z\|^r_A)=0
\end{eqnarray*}
for all $x,y,z\in A$. So
$$\delta([x,y,z])=[\delta(x),y,z]+[x,\delta(y),z]+[x,y,\delta(z)]$$
for all $x,y,z\in A$.

By the same argument as in the proof of Theorem 2.2, the uniqueness of $\delta$ is proved.
Thus the mapping $\delta$ is a unique $C^*$-ternary derivation satisfying
(\ref{ter deri near f r>3}).
\qed

\begin{thm} 
Let $r<1$ and $\theta$ be positive real numbers, and let $f:A\to A$ be a mapping satisfying
$(\ref{app add r>3 B=A})$, $(\ref{app ter deri r>3})$ and $f(0)=0$. Then there exists a
unique $C^*$-ternary
derivation $\delta:A\to A$ such that
\begin{equation}\|f(x)-\delta(x)\|_A\le\theta\,\frac{\,2+3^r}{\,3-3^r}\,\|x\|_A^r
\label{ter deri near f r<1}
\end{equation}
for all $x\in A$.
\end{thm}

\noindent{\it Proof.}
By the same argument as in the proof of Theorem 2.3, one can obtain a $\mathbb C$-linear
mapping $\delta:A\to B$ satisfying (\ref{ter deri near f r<1}).

The rest of the proof is similar to the proof of Theorem 4.1.
\qed

\begin{thm} 
Let $r>\frac{1}{3}$ and $\theta$ be positive real numbers, and let $f:A\to A$ be a mapping
satisfying $f(0)=0$ such that
\begin{eqnarray}
&& \|D_\mu f(x,y,z)\|_A\,\le\,\theta\cdot\|x\|^r_A \cdot\|y\|^r_A\cdot\|z\|^r_A,
\label{app add r>1/3 B=A}\\
&& \|f([x,y,z])-[f(x),y,z]-[x,f(y),z]-[x,y,f(z)]\|_A\nonumber\\
&& \qquad\qquad\qquad\quad\,\le\,\theta\cdot\|x\|^r_A\cdot\|y\|^r_A\cdot\|z\|^r_A
\label{app ter deri r>1/3}
\end{eqnarray}
for all $\mu\in{\mathbb T}^1$ and all $x,y,z\in A$. Then there exists a unique $C^*$-ternary
derivation $\delta:A\to A$ such that
\begin{equation}\|f(x)-\delta(x)\|_B\le\frac{3^r\theta}{27^r-3}\|x\|^{3r}_A
\label{ter deri near f r>1/3}\end{equation}
for all $x\in A$.
\end{thm}

\noindent{\it Proof.}
By the same argument as in the proof of Theorem 2.4, one can obtain a $\mathbb C$-linear
mapping $\delta:A\to B$ satisfying (\ref{ter deri near f r>1/3}).

The rest of the proof is similar to the proof of Theorem 4.1.
\qed

\begin{thm} 
Let $r<\frac{1}{3}$ and $\theta$ be positive real numbers, and let $f:A\to A$ be a mapping
satisfying $(\ref{app add r>1/3 B=A})$, $(\ref{app ter deri r>1/3})$ and $f(0)=0$. Then there
exists a unique $C^*$-ternary
derivation $\delta:A\to A$ such that
\begin{equation}\|f(x)-\delta(x)\|_A\le\frac{3^r\theta}{3-27^r}\|x\|^{3r}_A
\label{ter deri near f r<1/3}
\end{equation}
for all $x\in A$.
\end{thm}

\noindent{\it Proof.}
By the same argument as in the proof of Theorem 2.5, one can obtain a $\mathbb C$-linear
mapping $\delta:A\to B$ satisfying (\ref{ter deri near f r<1/3}).

The rest of the proof is similar to the proofs of Theorems 4.1.
\qed

{\footnotesize

\end{document}